# Forecasting the Mix of World Energy Needs by mid-21st Century

**Theodore Modis***

***This work has been published as a chapter in a Springer Nature book edited by Devezas, T., et al. with title "Global Challenges in Climate Change, Technological Foresight and Risks Assessment."***

* Theodore Modis is the founder of Growth Dynamics, an organization specializing in strategic forecasting and management consulting: http://www.growth-dynamics.com
Address reprint requests to: Theodore Modis, Via Selva 8, 6900 Massagno, Lugano, Switzerland. Tel. +41-91-9212054, E-mail: tmodis@yahoo.com

**Abstract**

The logistic function is used to forecast energy consumed worldwide. The logistic substitution model is used to describe the energy mix since 1965 presenting a picture significantly different from the one covering the previous 100 years. In the new picture the share of heavy pollutants, i.e. coal plus oil, keeps declining systematically in favor of natural gas and renewables (wind, geothermal, solar, biomass, and waste), the share of which grows rapidly. The shares of these three energy sources—coal+oil, natural gas, and renewables—are poised to reach around 30% each by mid-21st century. Nuclear and hydroelectric energy, both with rather stable shares, are responsible for the remaining 10%, which goes mostly to hydroelectric.

Zooming into the composition of renewables we find that today's dominant wind power is about to begin losing share to solar energy, which will overtake wind after 2024 and account for more than 90% of all renewables by mid-21st century, by which time geothermal, biomass, and other renewable energy sources will have dropped to insignificant levels.

Forecasts in exajoules are given for all energy sources up to 2050.

## 1. Introduction

This work updates a 3-year-old publication (Modis 2019). But the data analyzed here, provided by *BP Statistical Review of World Energy, July 2021*, have been restated in exajoules and give rise to significantly different results.

### 1.1. *Background*

The logistic function is generally suitable to describe natural-growth processes. Originally designed for species populations it is dotted with predictive power as demonstrated by the fact that no ecological niche has ever remained incomplete under *natural* conditions. In nature, deviations from logistic trajectories occur whenever there are mutations, natural disasters, or other such "unusual" events. In society, deviations from logistic trajectories may result from ill-conceived government decisions, wars, or major technological breakthroughs. Deviations meet with resistance and those resulting from ill-conceived decrees are generally short-lived permitting the logistic trajectory to regain its course toward completion.

Because the logistic function describes a natural law it is dotted with a capability to make more accurate longer-term forecasts than simple curve-fitting techniques. We will concentrate on *physical variables* because economic indicators like prices are a rather frivolous means of assigning lasting value and will not necessarily follow a natural-growth curve. Inflation and currency fluctuations due to speculation or politico-economic circumstances can have a large unpredictable effect on monetary indicators. Extreme swings have been observed. For example, Van Gogh died poor, although each of his paintings is worth a fortune today. The art he produced has not changed since his death; counted in dollars, however, it has increased tremendously. Therefore our analysis using logistics will not consider arguments intrinsically connected to prices such as energy return on investment (EROI).

Fisher and Pry have demonstrated that the logistic function is also suitable to describe competitive substitutions (Fisher and Pry 1971). It was in 1977 that Marchetti first used logistic substitution models to study the primary energy mix worldwide (Marchetti 1977). Later Nakicenovic implemented a software program using a generalized logistic substitution process (Nakicenovic 1979). The program facilitated long-range forecasts for the world energy consumption and yielded an elegant picture that became classic (Marchetti 1987). The implication was that a few parameters that described well the trajectories of the entry and exit of primary energies on the world stage for more than one hundred years would continue doing so in the long-term future. Since that time there have been many publications with updates of this picture (Ausubel et al. 1988), (Grubler 1990), (Nakicenovic et al. 1991), (Modis 2009), (Modis 2014).

As time went by however persisting deviations from the model began making their appearance. They were generally dismissed as “to be soon reabsorbed” by proponents of logistics. Work coming out from IIASA maintained an unchanged model as late as 2002 (Nakicenovic et al. 1995), (Nakicenovic 2002). However, Fisher and Pry had warned us of substitutions that may not proceed to completion due to “locked” market segments.

Devezas et al. formally addressed the important deviations from Marchetti’s original energy-substitution picture. They wrote, “an astonishing deviation … can be observed … the pattern was broken and replaced by relative flatness from the mid-1980s” (Devezas et al. 2008). In fact, upon more careful observation one can detect the beginning of such deviations as early as the mid-1960s. Devezas gives credit to Smil for first pointing out a significant deviation in the energy-substitution picture. In 2000 Smil proposed that Marchetti was wrong in concluding that the system’s dynamics cannot be influenced, and stated that: “After 1973, many forces began reshaping the system on a massive scale…” (Smil 2000). But it was Stewart who had first pointed out a flattening of the energy trajectories more than ten years earlier (Stewart 1989).

Arguments have been made for energy substitutions correlating with Kondratieff waves (Marchetti 1986), (Berry 1997). But there is evidence that these waves may now be deviating from their regular patterns (Modis 2017). Russian economists raise the possibility of even the end of this cyclical phenomenon altogether (Grinin et al. 2016). Could such evolutions have contributed to the energy deviations mentioned above?

Today it has become compelling to revisit those logistic forecasts of that elegant world-energy picture not only in order to check the validity of that model but also because long-term forecasts of energy needs always remain a topic of vital interest and discussion.

The analysis in this article assigns quantitative confidence levels on the uncertainties of forecasts whenever realistically feasible. The latest data available are used—up to July 2021—from the *BP Statistical Review of World Energy 2021*. Older data come from the *Statistical Abstract*, the Census Bureau of the U.S.

## 2. Worldwide Energy Consumption

The Census Bureau of the U.S. gives historical data on world energy consumption going back to 1860 and up to 1965. With more recent data from the *BP Statistical Review* we see in Figure 2.1 the yearly evolution of this variable up to the end of 2020. The data points generally follow a logistic trajectory albeit wiggling somewhat around it. In fact, one may want to discern a finer structure consisting of two or three smaller constituent logistics. However, it was decided to fit a single overall logistic to the entire data set (thick gray line) in order to obtain more reliable long-term forecasts. The Census Bureau data, which were expressed in tons of oil-equivalent were transformed into exajoules in order to be consistent with the *BP Statistical Review* data.

The data cover 62% of the logistic by the end of 2020. We estimate uncertainties on the parameters of the fitted logistic from look-up tables in the detailed Monte Carlo study by (Debecker and Modis 1994). The uncertainty on the final ceiling of the logistic is estimated as ±14% and the midpoint in 2005 has an uncertainty of ±6 years, both with 90% confidence level. The dotted lines in the figure delimit this uncertainty on the forecasted trajectory. It means that nine times out of ten future values should fall inside the band delimited by the dotted lines. The 2020 data point reflects the effect of COVID-19.

Table 1.1 gives forecasts and associated error estimates up to 2050 in exajoules.

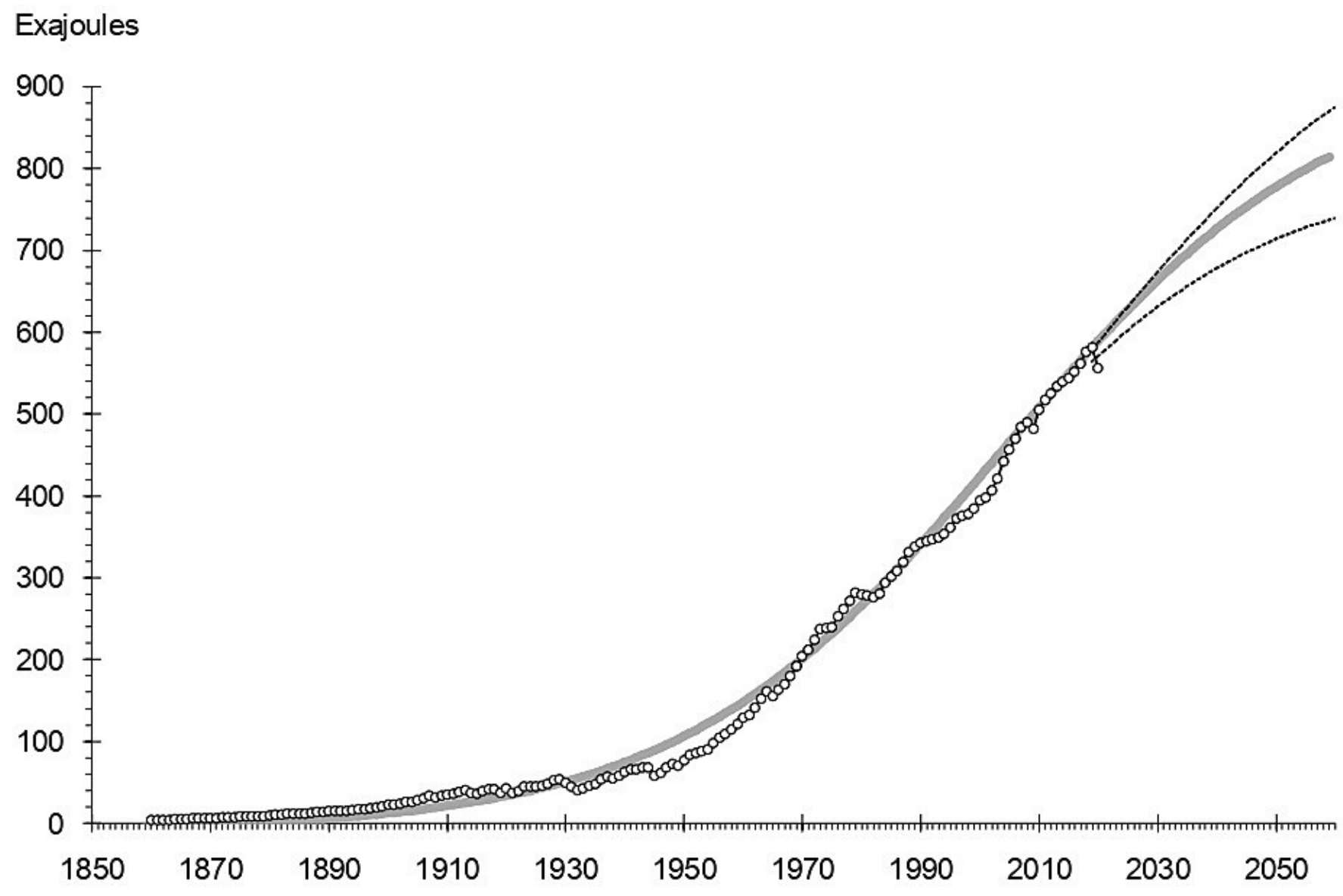

**Figure 2.1** Annual energy consumed worldwide (yearly data). The gray line is a logistic fit. The dotted lines delimit the 90% confidence level. The effect of COVID on the world economy shows on the point of 2020.

**Table 1.1 Forecasts and uncertainties for world energy consumption**

| | Exajoules | Error |
|---|---|---|
| 2022 | 597 | +1.4%<br>-2.3% |
| 2025 | 620 | +1.8%<br>-2.9% |
| 2030 | 656 | +2.6%<br>-4.0% |
| 2035 | 689 | +3.4%<br>-5.1% |
| 2040 | 720 | +4.3%<br>-6.2% |
| 2045 | 748 | +5.1%<br>-7.3% |
| 2050 | 773 | +5.9%<br>-8.3% |

## 3. The Primary Energy Mix

The classic graph on primary energy substitutions published by Marchetti in 1987 is reproduced in Figure 3.1 with the data updated to the end of 2020; the small circles designate updates. The graph shows the shares of primary energy sources plotted with a non-linear (logistic) vertical scale, which transforms S-curves into straight lines.

It becomes immediately evident that some of the deviations from the smooth trajectories of the model, introduced as early as the mid-1960s—notably on the coal and the natural gas trajectories—not only did not become reabsorbed later but became more pronounced. The elegant model description that did fair justice describing the data for more than one hundred years is unquestionably no longer valid. From the four primary energies considered only oil behaved according to the model and even that shows persistent deviations during the last fifteen years. It is worth pointing out that Marchetti back in 1987 had envisaged a future energy source that he called "Solfus"—representng a combination of solar energy and nuclear fusion—which he estimated would enter the world market with 1% share around the mid-2020s.

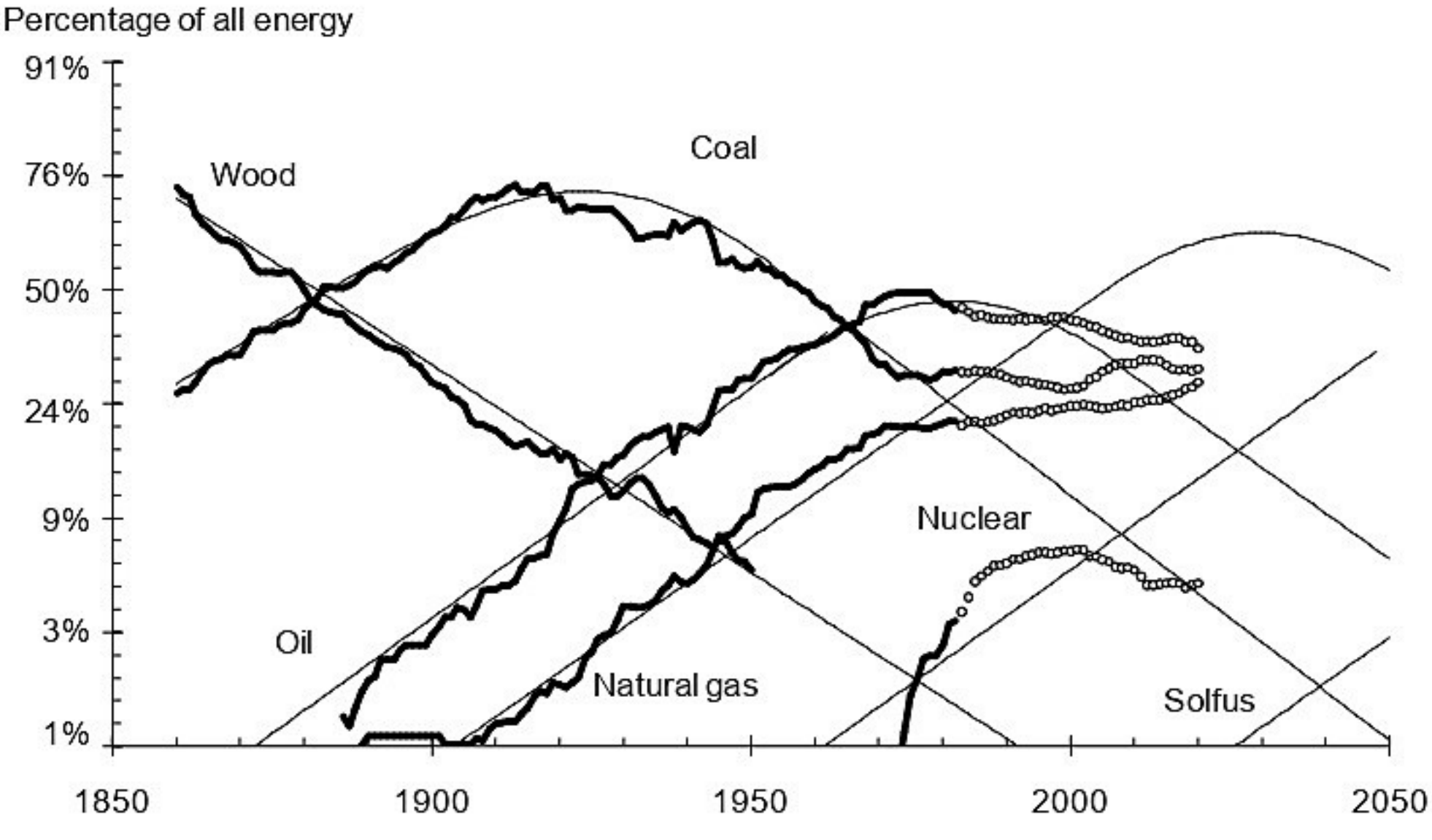

**Figure 3.1** Shares of primary energy sources. The black lines reproduce the picture published by Marchetti in 1987 (Marchetti 1987). The small circles are updates up to the end of 2020. The thin black lines represent the logistic substitution model. The vertical scale is non-linear (*logistic*) and transforms S-curves into straight lines. "Solfus" is a hypothetical future energy source coined by Marchetti, a combination of solar and nuclear fusion.

In Figure 3.1 renewable energy sources—comprising wind, geothermal, solar, biomass, and waste—are missing. This is because when Marchetti first put together the original graph renewable sources represented less than 1% share of the world market and did not make it into the graph. However, the renewable microniche has since become progressively more important and can no longer be neglected. Moreover, there is some complementarity between the coal and the oil trajectories; these two seem to form another microniche. Therefore, a new attempt is undertaken below to use the logistic substitution model to describe an enhanced set of five primary-energy sources with combinations that yield straight-line (logistic) behavior. We will zoom into the two microniches later. The historical window begins in 1965—when deviations first appeared—and for which time period there are data from only one source, the *BP Energy Review*, thus ensuring data consistency.

## 4. Five Competitors

In Figure 4.1 we see the evolution of the market shares of the five competitors; coal and oil are grouped together. The data trend in four of the five trajectories agrees with a straight line, which translates to logistic behavior (S-curve) in this graph with a logistic vertical scale. The trajectory of nuclear first goes over a peak, but then it also adopts a straight-line trend from 2012 onward. In the forecasting horizon—thin lines between 2020 and 2050—the share of renewables is calculated as 100% minus the other four shares, each one of which continues on a straight-line trend. The forecasted straight lines for nuclear and (coal + oil) have been determined from the data between 2012 and 2020.

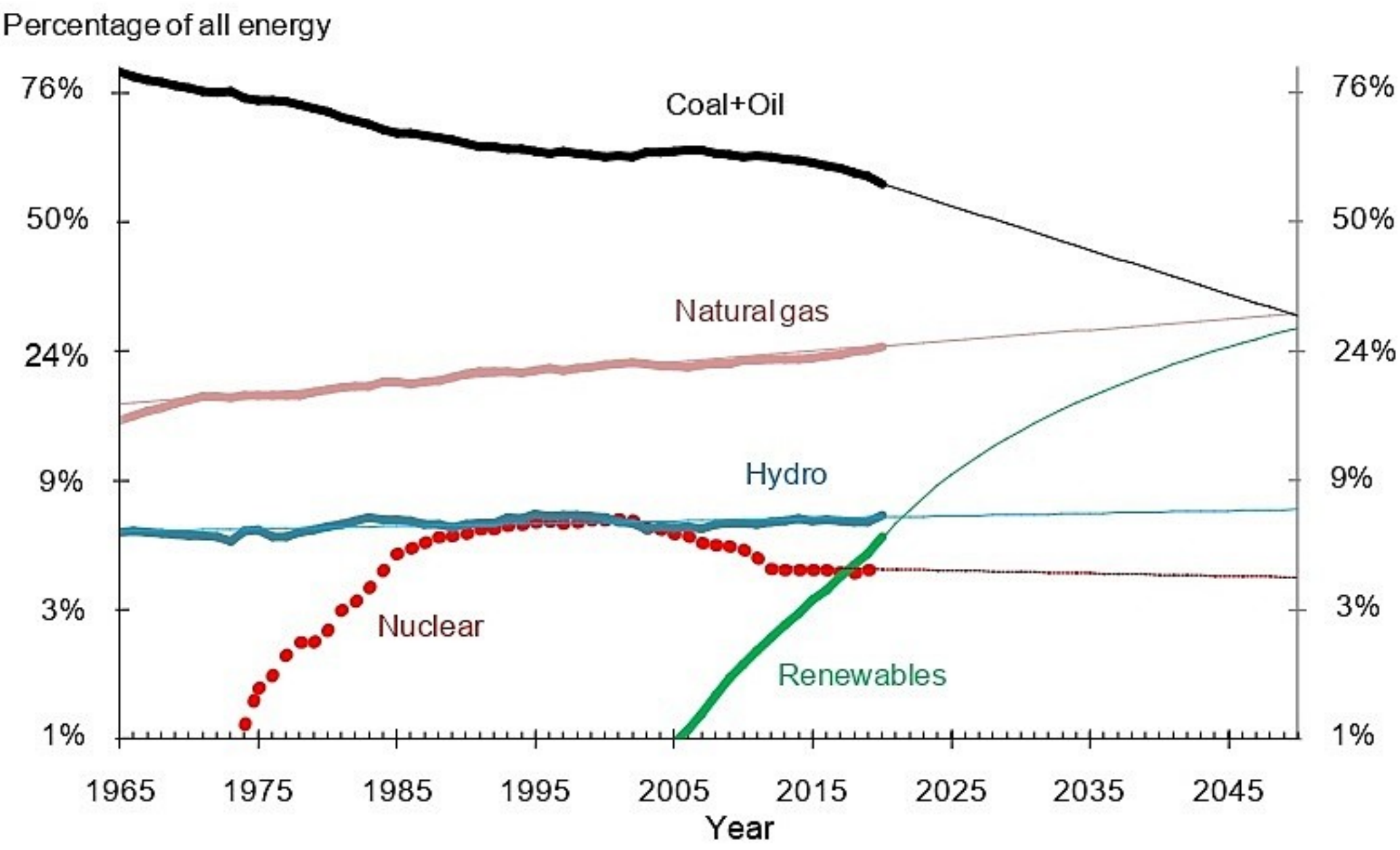

**Figure 4.1** Shares for the primary energy sources worldwide since 1965. The vertical scale is non-linear (*logistic*). The thick lines track annual data; the thin lines are forecasts by the logistic substitution model.

The forecasts for market shares from Figure 4.1 can be combined with the forecasts for the total energy from Table 1.1 to yield energy forecast in exajoules for each one of the five primary energy sources, see Table 4.1.

**Table 4.1 Forecasts for the five competitors**

| | Market Shares | | | | | Energy forecasts (exajoules) | | | | | |
|---|---|---|---|---|---|---|---|---|---|---|---|
| | Coal+ Oil | Nat. Gas | Nuclear | Hydro | Renew-ables | Total | Coal+ Oil | Nat. Gas | Nuclear | Hydro | Renew-ables |
| 2022 | 57% | 25% | 4% | 7% | 7% | 597 | 337 | 151 | 26 | 40 | 43 |
| 2025 | 54% | 26% | 4% | 7% | 10% | 620 | 332 | 160 | 27 | 42 | 59 |
| 2030 | 49% | 27% | 4% | 7% | 13% | 656 | 319 | 175 | 28 | 45 | 88 |
| 2035 | 44% | 28% | 4% | 7% | 17% | 689 | 302 | 191 | 29 | 48 | 120 |
| 2040 | 39% | 29% | 4% | 7% | 21% | 720 | 281 | 206 | 30 | 51 | 152 |
| 2045 | 34% | 30% | 4% | 7% | 25% | 748 | 258 | 221 | 31 | 53 | 185 |
| 2050 | 30% | 31% | 4% | 7% | 28% | 773 | 233 | 236 | 31 | 55 | 216 |

## 5. Two Microniches

### *5.1 Renewables*

In the group of renewables we will consider three members: wind, solar, and (geothermal + biomass + other.) Figure 5.1 presents a typical logistic-substitution picture. The share of (geothermal + biomass + other) is declining, the share of solar is growing, while the share of wind dominates but has stopped growing and is poised to enter a decline. Again, roughly straight-line trajectories reflect logistic behavior. The forecasted thin straight lines for solar and (geothermal + biomass + other) are calculated from the data of years 2012 to 2020. The forecasted trajectory for wind is calculated as 100% minus the other two straight lines. After 2023, when the share of wind has entered a declining straight-line trend, it is the share of solar that is now calculated as 100% minus the other two straight lines.

The forecasts for market shares from Figure 5.1 can be combined with the energy forecasts from Table 4.1 to give forecasts in exajoules for these three renewable-energy sources, see Table 5.1.

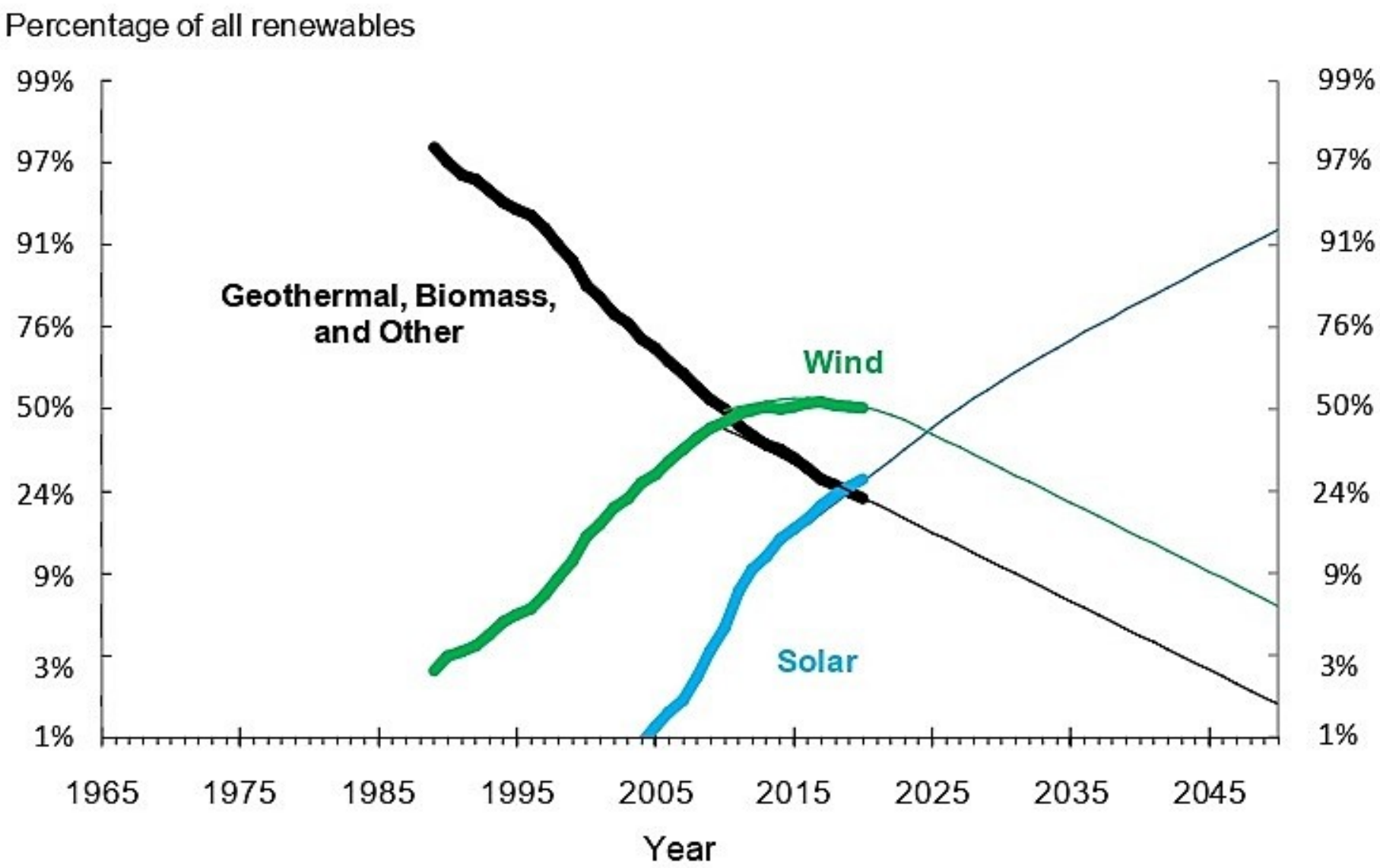

**Figure 5.1** Shares for (geothermal + biomass + other), wind, and solar in the renewables microniche. The vertical scale is non-linear (*logistic*). The thick lines are annual data; the thin lines are descriptions by the logistic substitution model.

**Table 5.1 Forecasts for renewable**

| | Market Shares | | | Energy forecasts (exajoules) | | | |
|---|---|---|---|---|---|---|---|
| | Wind | Solar | Rest | Total | Wind | Solar | Rest |
| 2022 | 48% | 33% | 19% | 43 | 20 | 14 | 8 |
| 2025 | 41% | 43% | 15% | 59 | 24 | 26 | 9 |
| 2030 | 30% | 60% | 10% | 88 | 27 | 53 | 9 |
| 2035 | 21% | 72% | 6% | 120 | 25 | 87 | 8 |
| 2040 | 14% | 82% | 4% | 152 | 22 | 125 | 6 |
| 2045 | 9% | 88% | 3% | 185 | 17 | 163 | 5 |
| 2050 | 6% | 93% | 2% | 216 | 13 | 200 | 3 |

*5.2 Coal + Oil*

In order to separate coal from oil we study the microniche Coal + Oil, see Figure 5.2. The two trajectories are obviously complementary but also rather flat. Straight lines are fitted on the last fourteen years of the data series. These are the logistic forecasts, indicated with thin lines in the figure, and showing a practically stationary evolution with oil dominating.

Once again, the forecasts for market shares from Figure 5.2 can be combined with the energy forecasts from Table 4.1 to give forecasts in exajoules for coal and oil separately, see Table 5.2.

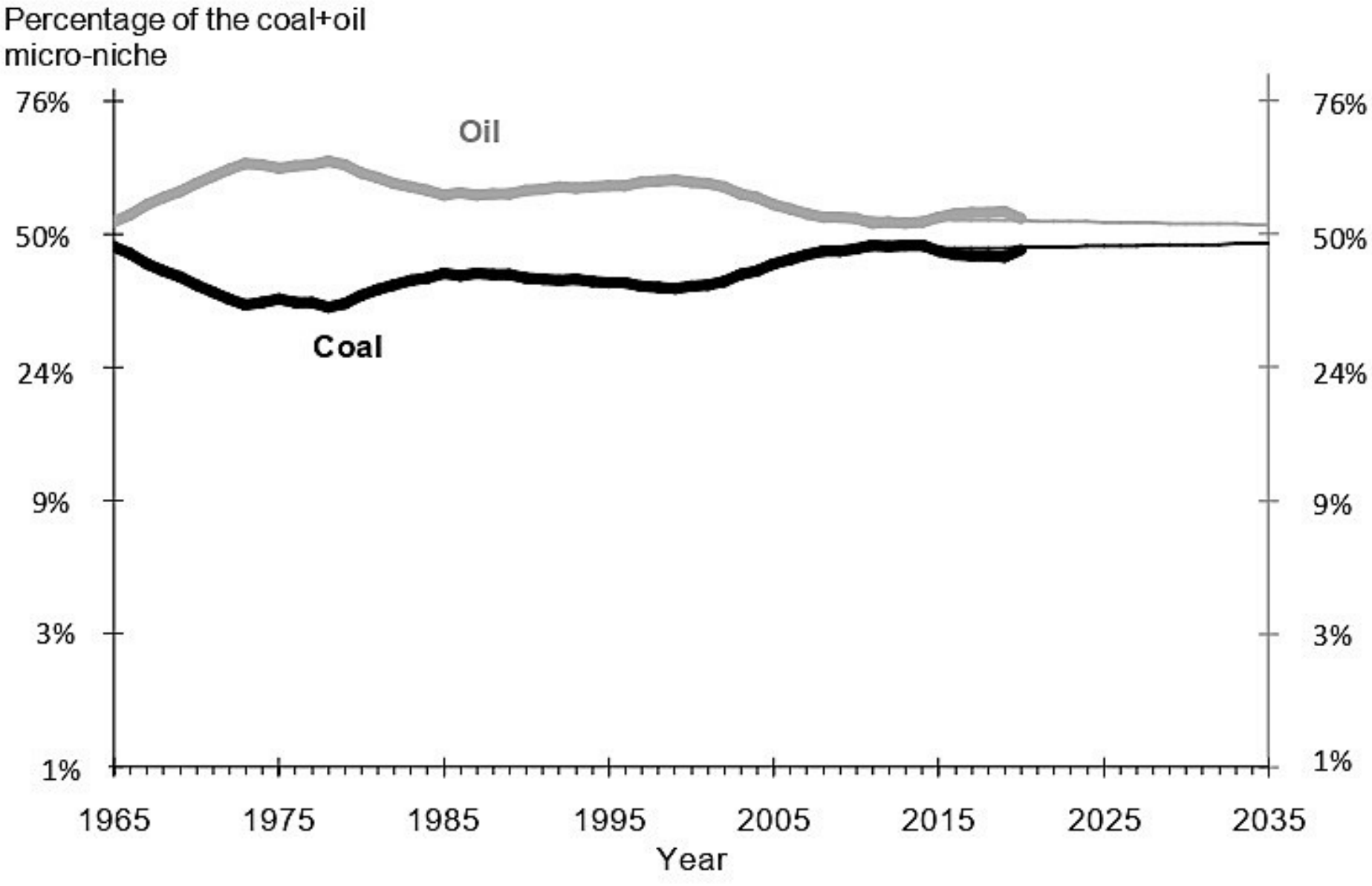

**Figure 5.2** Shares for oil and coal in the Oil + Coal microniche since 1965. The vertical scale is non-linear (*logistic*). The thick lines are annual data; the thin lines are forecasts by the logistic substitution model.

**Table 5.2 Forecasts for oil and coal**

| | Market Shares | | Energy forecasts (exajoules) | | |
|---|---|---|---|---|---|
| | Oil | Coal | Coal+Oil | Oil | Coal |
| 2022 | 53% | 47% | 337 | 178 | 159 |
| 2025 | 53% | 47% | 332 | 175 | 157 |
| 2030 | 52% | 48% | 319 | 167 | 152 |
| 2035 | 52% | 48% | 302 | 157 | 145 |
| 2040 | 52% | 49% | 281 | 145 | 139 |
| 2045 | 51% | 49% | 258 | 133 | 125 |
| 2050 | 51% | 49% | 233 | 119 | 114 |

## 6. Discussion

The overall energy consumption worldwide has been described by a logistic curve spanning the last 150 years. The forecast thus obtained indicates a slowing down of the rate of growth toward the mid-21$^{st}$ century.

The historical window—1965 to 2020—was studied in detail in order to take into account the rise of renewables and in order to better understand the deviations observed in the previous energy picture. It is only after grouping the shares of all renewable (except hydroelectric) together, and also oil plus coal together, that it becomes possible to use the generalized logistic substitution model successfully. This grouping into microniches also makes broader sense. Coal and oil are the worst pollutants, they are both on the decline, and they have trajectories to a large extent complementary, while the renewable niche presents a coherent picture of its own.

Devezas et al. addressed the same problem by grouping together oil with natural gas arguing that they have similar geological origins and locations, are extracted with similar technologies, often by the same commercial organization, and are often transported through pipes (Devezas et al., 2008). Their classification restores the validity of the substitution model but only up to the mid-1970s. From then onward they need to take into account energy efficiency calculations involving estimates for the world GNP. And they consider the specific future mix of renewables and nuclear energy as "uncertain."

Our choice to group oil with coal is based on their heavily polluting nature and the growing environmental awareness. This is amply demonstrated in Figure 4.1, which shows that their trajectory is unambiguously on the decline. The arguments by Devezas et al. about similarities in extraction and transport techniques between oil and natural gas are of secondary importance. These arguments concern technical issues, whereas the shift toward cleaner energies is a fundamental phenomenon on a worldwide scale.

Moreover, this grouping validates the substitution model well into the 21$^{st}$ century with no need to invoke efficiency arguments, which involve unreliable monetary indicators. Energy efficiency has been steadily growing over the past decades (probably following a logistic curve of its own) but its influence has already been taken into account in the calculation of the logistic forecasts for overall energy consumption.

In the emerging picture renewables are taking market share from (coal + oil), while natural gas keeps growing at the same slow rate. The three of them claim roughly 30% each of the overall energy market by 2050. Hydro and nuclear remain rather flat reaching 7% and 4% respectively by 2050. As for the composition of renewables, despite the dominant position of wind today with around 50% share of all renewables the rising star is solar, which is poised to overtake wind around 2024 and dominate with more than 90% share of all renewables by 2050, (other renewable energy sources will have dropped to negligible levels by then.) The uncertainties on the overall energy consumption forecasts—estimated via the look-up tables (Debecker and Modis 1994)—are not propagated down to the level of individual energy types because the forecasts on the shares carry additional uncertainties that cannot be reliably quantified.

The most striking highlight in this work is the emerging importance of solar energy as a primary-energy source. From Tables 4.1 and 5.1 we can calculate that solar energy will account for 26% of all energy consumed by 2050. This will not necessarily be all done using photovoltaic cells and other types of solar panels. New ways of harvesting sunlight are being developed using large concave mirrors to concentrate solar radiation for the dissociation of water and production of hydrogen via an integrated thermo-chemical reactor/receiver system, see for example the work led by A. G. Konstandopoulos (Agrafiotis et al. 2005).

The emerging importance of solar energy should not come as a surprise if we remind ourselves that all energy on earth—except for nuclear—comes from the sun anyway. The new realization is that it should be *direct* solar energy and not via wind, hydro, sea waves, oil, carbon, etc. Back in the 1980s Marchetti had envisioned a futuristic energy source that he called "solfus" a combination of solar and nuclear fusion to enter the world market by mid 2020s. He seems to have been on the right side, at least to some extent.

But nuclear fusion is not with us yet, and personally I am pessimistic. I first heard about it as an undergraduate in engineering school in the mid 1960s. We were told that research work was going on and industrial exploitation should be expected in twenty years. But in the mid-1980s I was at CERN and heard that a Tokamak at the École Polytechnique Fédérale de Lausanne (EPFL) would prove feasibility of a project that would lead to the production of fusion energy industrially in twenty years. But by the turn of the century, despite much progress in the meantime, the date was pushed back another twenty years. I recently read in the magazine of Nuclear Engineering International that "The Spherical Tokamak for Energy Production (STEP) project, as it is known, is being overseen by the UK Atomic Energy Authority (UKAEA), and aims to have an operational facility by 2040" (Nuclear Engineering International, 2021). The 20-year horizon seems to be an *invariant*, something very easy to forecast!

As mentioned in the introduction natural-growth processes—described by the logistic function—proceed to completion *under natural conditions*. This should not be interpreted as the forecasts made in this article will come true if nothing is done. Natural conditions for a logistic fit include all the kinds of things that took place during the historical window used for the fit. Actions by environmentalists, technological improvements, awareness of the public and the politicians, discovery of new reserves, and economic development worldwide have all been folded in while fitting the logistic function. In the future policy makers will not have a free hand to do whatever they wish. They must continue listening to the growing voices of the public about climate change and the environment, and responding proportionally to the loudness of these voices, just as they have been doing in the past. The continuation of all well-established process on their *natural-growth* trajectories will ensure that the forecasts made in this paper come true.

Deviations could be caused only by the kind of major events never seen before tantamount to a "mutation" in the evolution of a species. But new breakthroughs in technologies such as artificial intelligence, bioengineering, or nanotechnology would not impact the evolution of most trajectories forecasted above because the beginning of these technologies has already been folded in the evolution of the data during recent years, and also we have seen this kind of technologies emerge regularly during the historical window. On the other hand, unseen-yet events of greater importance, e.g. major wars or the mastering of thermonuclear fusion would certainly interrupt/modify the forecasted trajectories.

**Endnote**

**Theodore Modis** is a physicist, strategist, futurist, and international consultant. He is author/co-author to over one hundred articles in scientific and business journals and ten books. He has on occasion taught at Columbia University, the University of Geneva, at business schools INSEAD and IMD, and at the leadership school DUXX, in Monterrey, Mexico. He is the founder of Growth Dynamics, an organization specializing in strategic forecasting and management consulting: http://www.growth-dynamics.com.